# Dual Fano and Lorentzian line profile poperties of autoionizing states

B. Tu[1], J. Xiao[1], K. Yao[1], Y. Shen[1], Y. Yang[1], D. Lu[1], W. X. Li[1], M. L. Qiu[1], X. Wang[1], C. Y. Chen[1], Y. Q. Fu[1], B. Wei[1], C. Zheng[1], L. Y. Huang[1], R. Hutton[1], and Y. Zou[1*]

[1] Shanghai EBIT Laboratory, Institute of Modern Physics, Fudan University, and the Key Laboratory of Applied Ion Beam Physics, Chinese Ministry of Education, China

*Correspondence to: zouym@fudan.edu.cn

Ott *et al.* (*Science* (**340**, 716 (2013))) successfully transferred Fano profile into Lorentzian lineshape using an intense infrared laser, after excitation of autoionizing states in helium by attosecond XUV pulse. This is a very important step forward of quantum phase control. However, here we show experimentally that an autoionizing state can have both Fano and Lorentzian behavior naturally, depending on the process involved. This study utilized the inverse process of photon absorption ionization, i.e. electron ion recombination with photon emission, making sure the resonant autoionizing state is not perturbed by the laser fields. Our result implies that excitation of the state through different paths can lead to different Fano profiles for the same resonant state. This allows more options for the combination of laser fields and lead to more opportunities for quantum phase control. Our result also indicates the breakdown of the classical two step picture for dielectronic recombination.

Photon absorption spectroscopy is a powerful tool for uncovering the structure of atoms, molecules and solids. Asymmetric Fano absorption line shapes emerge when discrete excited states are coupled to the continuum [2, 3]. Such asymmetric resonances are ubiquitous in essentially all fields, through nuclear [4], atomic [5-8], molecular [9] to solid-state physics [10-13]. Coupling (or configuration interaction) of discrete and continuum states is a problem of electron-electron interaction, or electron-electromagnetic field interaction. The advanced laser technology nowadays allows scientists to simulate, modify, or even cancel some of the interactions, as was done in the work of Ott *et al.* [1], and finally could lead to many important applications [14-17], especially in the field of quantum control.

Electron ion recombinations, both resonant and non-resonant, are inverse processes of atomic photon absorption ionization. Dielectronic recombination (DR), as the main process of resonant recombination, is considered as a two-step process. In the first step a free electron is captured by an ion, and a bound electron in the ion is promoted, forming an intermediate doubly or even multiply excited state sitting above the ionization threshold (this kind of state is also called an autoionizing state). In the second step, photons are emitted to reduce the energy of the recombined ion to below its ionization limit, so as to stabilize this recombination. The first step endows DR the resonant character and determines the resonant energy, *i.e.* only when the energy released from the capture of the free electron matches the excitation energy of the promoted electron can the DR process happen. The second step can lead the recombined ion to different final states according to the transition branching ratio. In contract to this, in the non resonant recombination, also named as radiative recombination (RR), a free electron is captured by an ion and the excess energy is released directly by emitting a photon (see figure 1 a). DR and RR are also important in high temperature plasmas, like celestial or fusion plasmas. They affect the charge state balance, degrade the plasma temperature, and produce lots of satellite lines some of which are very useful in plasma diagnostics. Much work has been done to study these processes (see refs 18-22 and references therein), but only very few have observed interference between the DR and RR [23-25], due to the fact that most studies were for the cases where RR was much weaker than DR, and the interference was negligible.

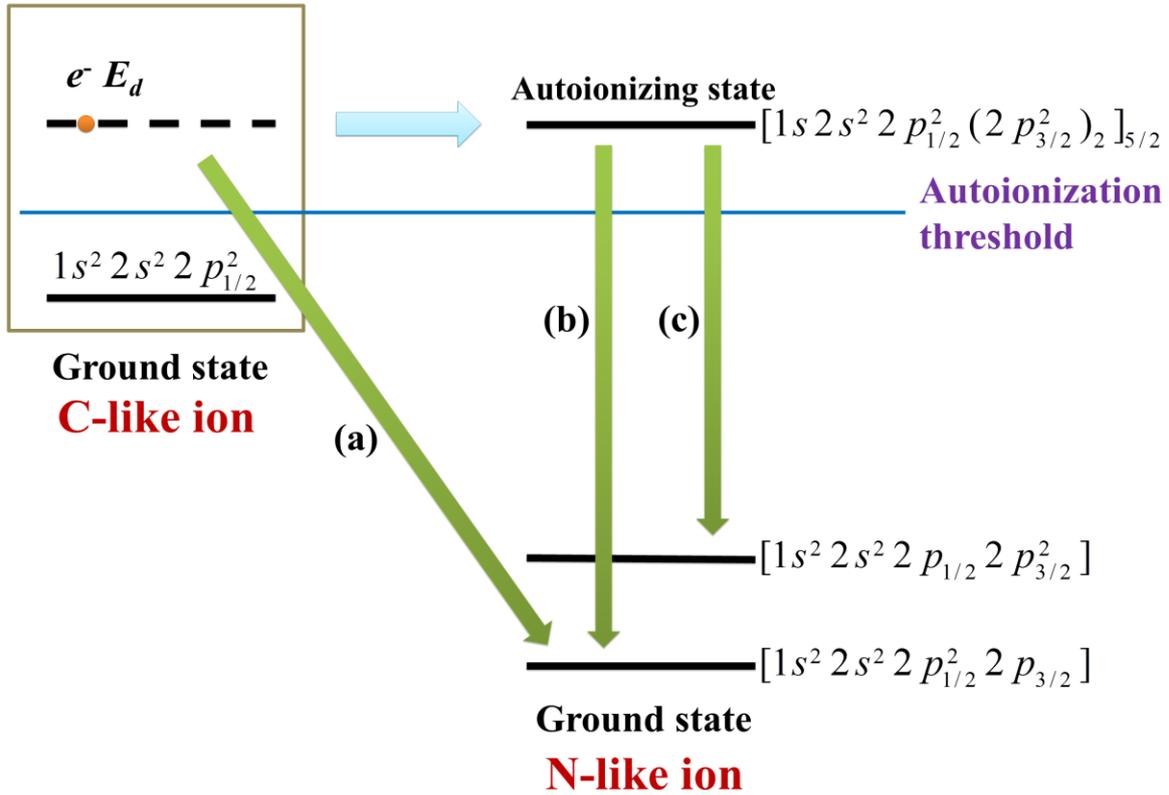

**Fig. 1. Energy level diagram of electron ion recombination for C-like ions.** (a) shows the radiative recombination of an electron with a C-like ion in the ground state, to form the N-like state $[1s^2 2s^2 2p_{1/2}^2 2p_{3/2}]$, (b) and (c) shows the dielectronic recombination of an electron with a C-like ion in the ground state, to form N-like state of $[1s^2 2s^2 2p_{1/2}^2 2p_{3/2}]$ and $[1s^2 2s^2 2p_{1/2} 2p_{3/2}^2]$ respectively, through the same autoionizing state of $[1s2s^2 2p_{1/2}^2(2p_{3/2}^2)_2]_{5/2}$.

In order to study the variation of the Fano profile (consequently the Fano parameter) of a given autoionizing state, we selected the DR resonances which go through an intermediate state with decay channels to both the final states with no excited electrons (figure 1 b) and the final states with two excited electrons (figure 1 c). In the former case, both dielectronic and radiative recombination will occur, interference and hence the asymmetric line profile should be expected. In the latter case, only dielectronic recombination (so no interference) will take place, because radiative recombination of an electron with an ion in the ground state can not end up in a final state with more than one electron in excited state. In detail, we studied the DR processes of C-like tungsten (W) ions, through the same intermediate state of $[1s2s^2\,2p_{1/2}^2\,(2p_{3/2}^2)_2]_{5/2}$ to two different final states, $[1s^2\,2s^2\,2p_{1/2}^2 2p_{3/2}]$ and $[1s^2\,2s^2\,2p_{1/2}2p_{3/2}^2]$.

The experiment was done at the Shanghai electron beam ion trap [26] (Shanghai EBIT). The setup and method in this work were similar to those described in our previous paper [20, 27, 28] for DR studies of Xe ions. During the experiment, $W(CO)_6$ was continuously injected into the EBIT, and a plasma with ions from W, C and O was produced in the trap region. The electron energy was adjusted as shown in figure 2, to produce the right charge states of W ions, and to scan through the DR resonances of interest. The event mode data acquisition system was triggered by an arrival x ray photon (from the recombination event), and the photon energy and the corresponding electron energy were simultaneously recorded as shown in figure 3. In figure 3, the resonant peaks are from $KL_3L_3$ DR events, in which a free electron was captured to $L_3$ at the same time a $K$ shell electron in the ion was promoted to $L_3$ ($L_3$ means the $2p_{3/2}$ orbit), of Be-, B-, C-, N-, and O-like W ions. The events on the clear diagonal band correspond to the RR processes to the final states of principal quantum number $n=2$, and the total angular momentum $J=3/2$ of these ions.

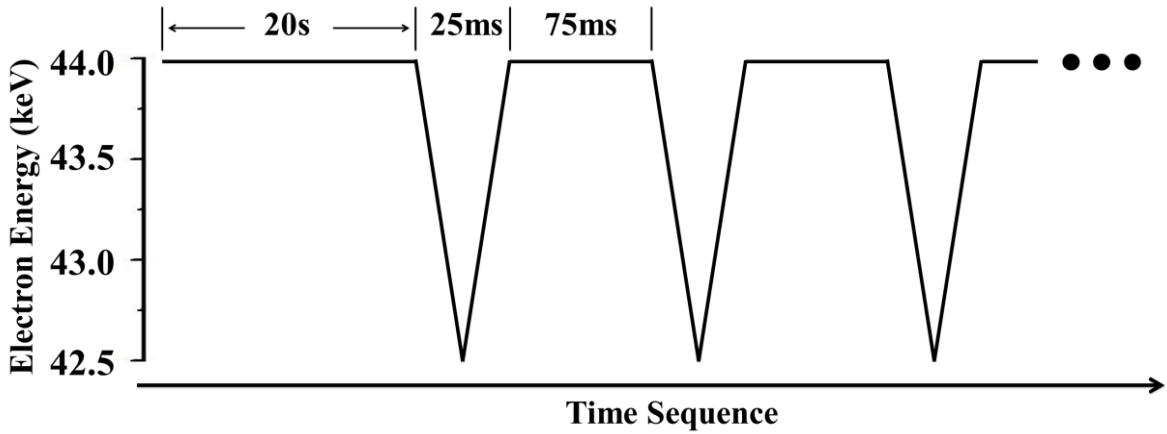

**Fig. 2. Time sequence for electron beam energy adjustment in the present experiment.**

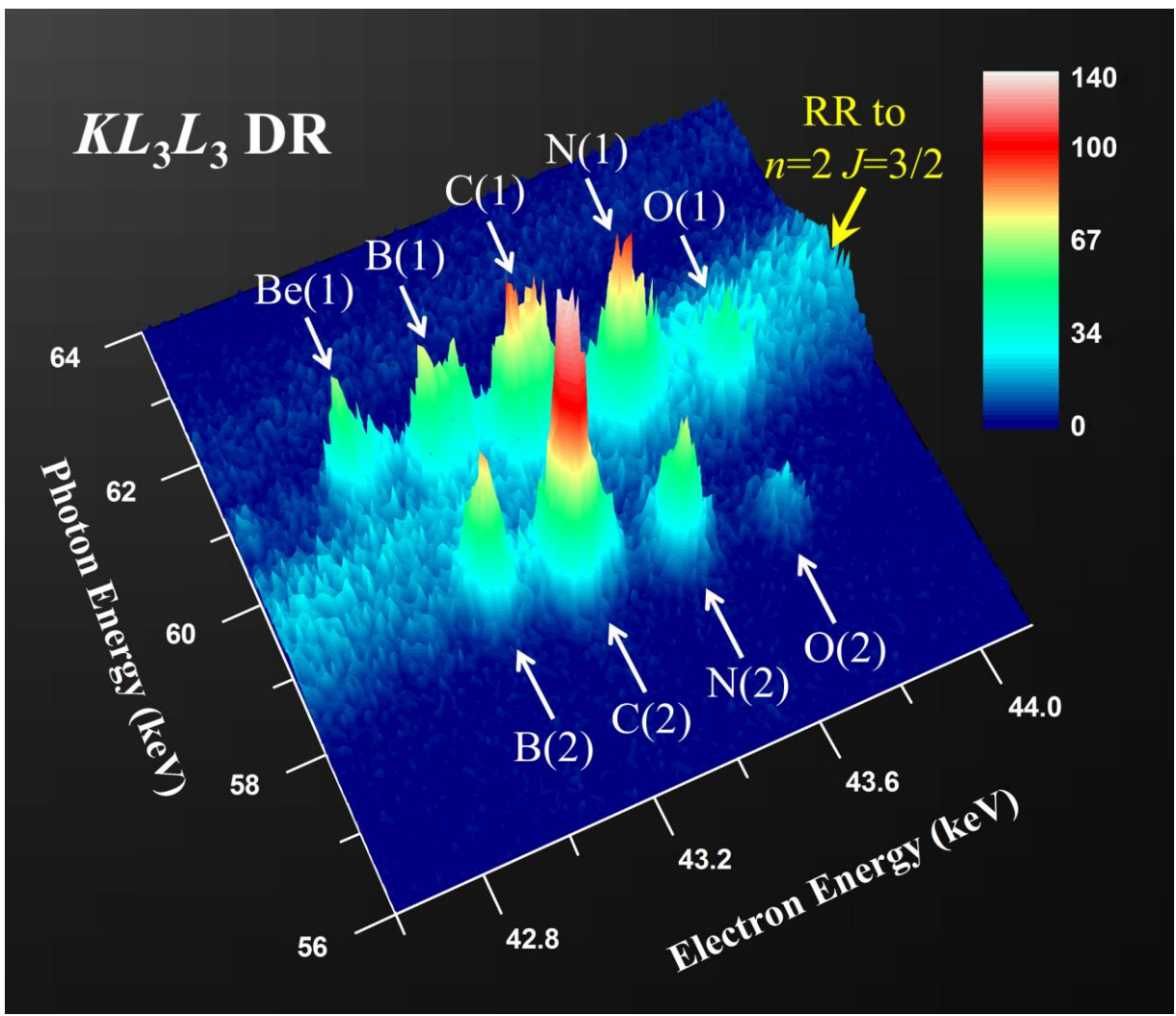

**Fig. 3. Three dimensional spectrum for the electron ion recombination of W ions.** The peaks are from

$KL_3L_3$ DR events of Be-, B-, C-, N- and O-like W ions. The events on the diagonal band correspond to the RR processes to the final states of principal quantum number $n=2$, and the total angular momentum of $J=3/2$ of these ions. The events in C(1) are from the DR processes through the autoionizing states of $[1s2s^2 2p^2_{1/2}(2p^2_{3/2})_2]_{5/2}$, $[1s2s^2 2p^2_{1/2}(2p^2_{3/2})_2]_{3/2}$, and $[1s2s^2 2p^2_{1/2}(2p^2_{3/2})_0]_{1/2}$, to the final state of $[1s^2 2s^2 2p^2_{1/2}2p_{3/2}]_{3/2}$. Those in C(2) are through the same intermediate states, but to the final states with two excited electrons, $[1s^2 2s^2 2p_{1/2}(2p^2_{3/2})_2]_{3/2}$, $[1s^2 2s^2 2p_{1/2}(2p^2_{3/2})_2]_{5/2}$, and $[1s^2 2s^2 2p_{1/2}(2p^2_{3/2})_0]_{1/2}$. The last final state is only from the $[1s2s^2 2p^2_{1/2}(2p^2_{3/2})_0]_{1/2}$.

The events marked as C(1) in figure 3 are from the DR processes through autoionizing states of $[1s2s^2 2p^2_{1/2}(2p^2_{3/2})_2]_{5/2}$, $[1s2s^2 2p^2_{1/2}(2p^2_{3/2})_2]_{3/2}$, and $[1s2s^2 2p^2_{1/2}(2p^2_{3/2})_0]_{1/2}$, to the final state of $[1s^2 2s^2 2p^2_{1/2}2p_{3/2}]_{3/2}$. Those designated as C(2) are through the same intermediate states, but to the final states with two excited electrons, $[1s^2 2s^2 2p_{1/2}(2p^2_{3/2})_2]_{3/2}$, $[1s^2 2s^2 2p_{1/2}(2p^2_{3/2})_2]_{5/2}$, and $[1s^2 2s^2 2p_{1/2}(2p^2_{3/2})_0]_{1/2}$. The last final state is only from the $[1s2s^2 2p^2_{1/2}(2p^2_{3/2})_0]_{1/2}$. The events in C(1) and C(2) were then picked out along the $n=2$, $J=3/2$ RR band and projected on to the electron energy axis, as was done in ref. 24, to produce excitation functions of electron ion recombination for the two cases, shown in figure 4a and 4b, respectively. Figure 4a shows asymmetric line profiles, 4b does not. It is obvious as C(1) sits on the RR band (so interference occurs), but C(2) not. We made a least square fitting for the excitation function in figure 4b, employing a Lorentzian Gaussian convoluted profile, to account for the small electron energy spread in the EBIT. The obtained resonant energies and the width of the Gaussian distribution were then used as fixed parameters, to fit the data in figure 4a, utilizing a Fano Gaussian convoluted profile [29], as in this case interference is involved. The

resonances for $[1s2s^2\,2p^2_{1/2}(2p^2_{3/2})_2]_{3/2}$ and $[1s2s^2\,2p^2_{1/2}(2p^2_{3/2})_0]_{1/2}$ are too close to resolve, and they are not the subjects for this study. The resonance for $[1s2s^2\,2p^2_{1/2}(2p^2_{3/2})_2]_{5/2}$ is just far enough from the un-resolvable ones, to obtain reliable results. Our data analysis led to a Fano parameter of 4.4 (0.4) for the interference channel through this autoionizing state. This agrees very well with our theoretical result of 4.3, based on relativistic configuration interaction method, using the Flexible Atomic Code [30]. Comparing the peak positions of the interference and non-interference channel in figure 4a and 4b (for clearer, see figure 4c and 4d), the phase shift in the energy domain caused by the interference was obtained for the first time in electron ion recombinations, and the value is 9.8 (0.8) eV.

The phase shift in the energy domain is actually a shift in resonant energy. In the classical picture, the resonant energy should be decided when the first step of the DR is taken. But our result shows the DR process involving the interference has its resonance pushed to higher energy relative to the non interference case, as if it knows already which final state will go before the first step happens. This indicates the breakdown of the classical two step picture for the dielectronic recombination process.

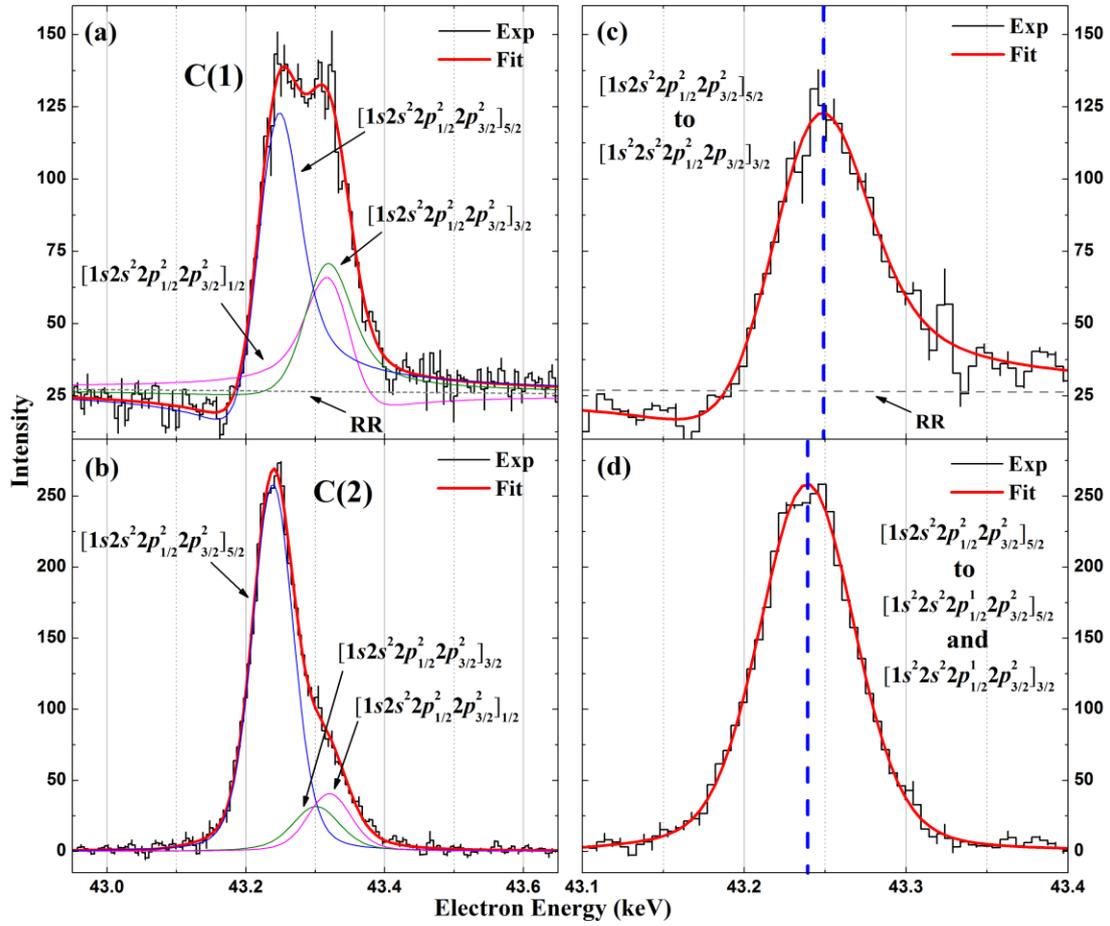

Fig. 4. The excitation functions of the KL$_3$L$_3$ DR spectra of C-like W ions through the interference channel (a) and non-interference channel (b). The intermediate autoionizing states are labeled for each resonance in (a) and (b). (a) shows an asymmetric line profile, which indicates the occurrence of the interference between the DR and RR. (b) shows a symmetric line shape, which implies that no interference is involved. (c) and (d) show the enlarged peak of the DR resonance through the autoionizing state of $[1s2s^2\,2p^2_{1/2}(2p^2_{3/2})_2]_{5/2}$ for the interference and non-interference channel, respectively, after subtracted the contribution from the two un-resolvable resonances, $[1s2s^2\,2p^2_{1/2}(2p^2_{3/2})_2]_{3/2}$ and $[1s2s^2\,2p^2_{1/2}(2p^2_{3/2})_0]_{1/2}$, according to the least square fitting result. The phase shift in the energy domain caused by the interference can be seen clearly in (c), comparing the peak position in (d).

The autoionizing states involved in the two cases are the same state, *i.e.* $[1s2s^2\,2p^2_{1/2}(2p^2_{3/2})_2]_{5/2}$, so should be described by the same wave function and hence the same mixing of discrete and continuum configurations. The very different Fano parameters for the two cases from our experiment, 4.4 (0.4) and infinite, indicates that an autoionizing state can have both Fano and Lorentzian behavior naturally, depending on the process involved.

Although coupling of a discrete state with the continuum is necessary for a Fano profile (without external disturbance), it does not mean that stronger coupling would lead to stronger Fano asymmetry. The Fano parameter is proportional to the relative strength of the DR over RR in electron ion recombination, or proportional to the relative strength of the indirect (resonant) ionization over the direct ionization in photon absorption. It is a measure of the degree of interference. The strongest asymmetric line profile, or the strongest interference, would occur when the strengths of the two interfering processes are about equal. With the absence of either partner, the interference will not occur, corresponding to the case of an infinite or a zero Fano parameter, and hence a vanishing Fano profile. But stronger coupling does mean stronger mixing of the continuum into the discrete, resulting from larger deviation from hydrogenic field for the electrons, which is usually seen in the cases where the two excited electrons are in close lying quantum states. In the work by Ott *et al.* [1], they could successfully turn the Fano profile to Lorentzian for $sp_{2n+}$ states with *n* above 3 for He, at the laser intensity of $2.0 \times 10^{12}$ W/cm$^2$, but not for 2s2p and $sp_{23+}$. It could well be the reason that the laser intensity is not enough to compensate the effect of stronger coupling of the continuum to the discrete for the 2s2p and $sp_{23+}$ cases. We could also expect that for $sp_{2n+}$ with very high *n*, the absorption line shape would approach Lorentzian even without the laser compensation, because in this case the wave function could be very well described by hydrogenic function, *i.e.* almost no interaction with continuum state. For the application to quantum control using Fano/Lorentzian line profile of atomic states, we suggest to chose the resonant states with medium or relatively weak coupling between the discrete states and the continuum for easier implementation of laser field modification of the coupling, and to chose the process with similar strengths of both channels to get the highest dynamic range of the interference.

**Acknowledgements:** The authors thank Dr. Sven Huldt and Dr. Tomas Brage for helpful comments. This work was supported by the Shanghai Leading Academic Discipline Project No. B107，and by the Shanghai Science and Technology commission under grant No. 13ZR1451500.